\documentclass[11pt]{article}
\usepackage{amsmath}
\usepackage{graphicx,psfrag,epsf}
\usepackage{enumerate}
\usepackage{url}
\usepackage{pdfsync}
\usepackage{amsmath}
\usepackage{amssymb}
\usepackage{graphicx}
\usepackage[latin1]{inputenc}
\usepackage{amsthm}
\usepackage[english]{babel}
{\catcode `\@=11 \global\let\AddToReset=\@addtoreset}
\AddToReset{equation}{section}

\newtheorem{propo}{Proposition}[section]
\newtheorem{@definition}{\sc Definition}[section]

\newtheorem{@remark}{\sc Remark}[section]

\newtheorem{@example}{\sc Example}[section]

\newtheorem{pf}{Proof}

\newcommand{\beqn}{\begin{displaymath}}
\newcommand{\eeqn}{\end{displaymath}}
\newcommand{\beq}{\begin{equation}}  
\newcommand{\eeq}{\end{equation}}

\def\mathsf{\bf}
\def\N{\mathbb{N}}

\def\R{\mathbb{R}}

\def\E{\mathrm E}

\def\text{\mbox}

\def\1{{\bf 1}}

\newcommand{\var}{\mbox{Var}}
\newcommand{\cov}{\mbox{Cov}}

\def\limitet{\renewcommand{\arraystretch}{0.5}
\begin{array}[t]{c}
\stackrel{}{\longrightarrow} \\
{\scriptstyle t\rightarrow\infty}
\end{array}\renewcommand{\arraystretch}{1}}

\def\limitet0{\renewcommand{\arraystretch}{0.5}
\begin{array}[t]{c}
\stackrel{}{\longrightarrow} \\
{\scriptstyle t\rightarrow 0}
\end{array}\renewcommand{\arraystretch}{1}}

\begin{document}

%

\title{
\begin{center}
New approximation for GARCH parameters estimate \\
{Yakoub BOULAROUK$^{1}$ , Nasr-eddine HAMRI$^{1}$} \\
$^{1}$  Institute of Science and Technology, Melilab laboratory, University Center of Mila, Algeria
\end{center}
}
\maketitle

{\bf Abstract}:    This paper presents a new approach  for the optimization of GARCH parameters estimation. Firstly, we propose a method for the localization of the maximum. Thereafter, using the methods of least squares, we make a local approximation for the projection of the likelihood function curve on two dimensional planes  by a polynomial of order two which will be used to calculate an estimation of the maximum.


\section{Introduction}
The modeling of time series is applied today in fields as diverse (econometrics, medicine or demographics....). As it is a crucial step in the study of time series, it has undergone a great evolution during the last fifty years and several models of representation have been proposed.\\
In 1982, Engle \cite {engel} proposed the conditionally heteroskedastic autoregressive (ARCH) model, which allowed the conditional variance to change as a function of past errors over time, while leaving the variance unconditional constant. This model has already proved useful in the modeling of several phenomenas. In Engle \cite {engel,engel83} and Engle and Kraft \cite {enrk}, models for the inflation rate are constructed recognizing that the uncertainty of inflation tends to change over time. In 1984, Weiss \cite {W2} considered the ARMA models with ARCH errors. He used these models for the modeling of sixteen US microeconomic time series.\\
In 1986, Bollerslev \cite {bol} proposed a generalization of these ARCH models to the Generalized AutoRegressive Conditional Heteroskedasticity (GARCH) models, whose the variance in the present depends on its past and the process past. As it studied the conditions of
stationarity and the structure of the autocorrelations for this class of models. \\
The GARCH family contains a number of parameters which must be estimated on actual data for empirical applications. The estimation of the parameters returns to calculate the maximum of the log-likelihood function which is non-linear, this leads to the use of non-linear optimization methods. Several optimization methods are used, including The Nelder-Mead method, BGFS method and others :
\begin{itemize}
  \item Nelder and Mead \cite{NeW} introduced their method using only

   the likelihood function values. Altough their method is relatively slow, it's robust and leads to results for the non differentiable functions. In their turn Fletcher and Reeves \cite{Fle} have introduced the method of the conjugate gradient which does not store matrix.
  \item In 1970 Broyden \cite{boyd}, Fletcher \cite{Flet}, Goldfarb \cite{gold} and Shanno \cite{sha} have published simultaneously the method entitled quasi-Newton (noted BFGS), which uses the values and the gradients of function to construct an image of the surface to optimize.
  \item In 1990 David M. Gay \cite{DAV} have published a technical report in which he proposed a method intitiled optimization using PORT routines (noted OPR).
\end{itemize}
It is common knowledge among practitioners that the GARCH parameters are numerically difficult to estimate in empirical applications. The existant numerical algorithm can easily fail, or converge to erratic solutions. Therefore, the resulting fitted parameters must be examined with a healthy dose of scepticism. In this work, we propose a new algorithm in which we exploit the asymptotic  convexity of the likelihood function. We will  develop a method to locate the maximum value instead of using the confidence intervals resulting from asymptotic normality of the estimated parameters thereafter and as in [2], we approach the likelihood function by a quadratic form that we use to calculate an approximation of the maximum.
\section{The GARCH model and its likelihood function}
The GARCH$(p,q)$ process was introduced by Bollerslev \cite{bol} as solution for the system of equations:
\begin{equation}\label{garch}
\left\{
  \begin{array}{ll}
& X_t= \sigma_t \cdot \varepsilon_t,\\
&\sigma_t^2=\omega+\Sigma_{i=1}^{p}\alpha_i X_{t-i}^2+\Sigma_{j=1}^{q}\beta_j \sigma_{t-j}^2.\\
  \end{array}
\right.
\end{equation}
 Let's note $\theta=(\omega,\alpha_1,...,\alpha_{p},\beta_1,...,\beta_{q})\in \R^{p+q+1}$ the vector of unknown parameters, with $\omega>0$, $\alpha_i\geq0$ for $i=1,p-1$, $\beta_j\geq0$ for $j=1,q-1$, $\alpha_{p}, \beta_{q}$ strictly positive. The $(\varepsilon_t)_{t\geq0}$ is a sequence of normal random independent and identically distributed satisfying the standard assumptions $\E(\varepsilon_t)=0 $ and $ \var(\varepsilon_t)=1$. The ARCH$(q)$ is an GARCH$(0,q)$\\
The conditional likelihood of $X$ expresses as, up to an additional constant,
\begin{eqnarray*}
  L_n(\theta)  &=& -\frac{1}{2}\sum^n_{t=1} q_t(\theta). \\
  \mbox{with} & & q_t(\theta)=log (\sigma_t^2) + \sigma_t^{-2} X_t^2
\end{eqnarray*}
The quasi-likelihood $\widehat{L}_n$ is obtained by plugging in $L_n$ the approximations $ \widehat{\sigma}_t^2=\sigma_t^2(X_t,...,X_1,u_n)$, where $u = (u_n)_{n\in \N}$ is different of zero only for finitely many $n\in \N$,
\begin{eqnarray*}
  \widehat{L}_n(\theta)  &=& -\frac{1}{2}\sum^n_{t=1} \widehat{q}_t(\theta). \\
  \mbox{with} & & \widehat{q}_t(\theta)=log (\widehat{\sigma}_t^2) + \widehat{\sigma}_t^{-2} X_t^2
\end{eqnarray*}
Remark that unobserved values $(X_t , t\leq 0)$ have to be fixed a priori equal to $(u_n)_{n\in N}$ in the quasi-likelihood $\widehat{L}_n$. In the next proposition, we give a necessary and sufficient condition for the process stationarity.
\begin{propo}\label{statio}
The GARCH( $p,q$) process as defined in (\ref{garch}) is wide-sense stationary with $E(\varepsilon_t)=0$, $\var(\varepsilon_t)= \alpha_0(\Sigma_{i=1}^{p} \alpha_i + \Sigma_{i=1}^{q}\beta_i)^{-1}$ and $\cov(\varepsilon_t, \varepsilon_s)= 0$ for $t\neq s$ if and only if
\begin{equation}\label{stat}
\Sigma_{i=1}^{p} \alpha_i + \Sigma_{i=1}^{q}\beta_i<1
\end{equation}
\end{propo}
\begin{pf} This result is proved by Bollerslev in \cite{bol}. \end{pf}
\begin{rm}
The condition (\ref{stat}) implies that $0\leq\alpha_i<1, \forall i=1,p$ and $0\leq \beta_j<1, \forall j=1,q$.
\end{rm}
Else, we define the stationarity set
\begin{equation}\label{stset}
  \Theta_0=\{\theta\in \R^{p+q+1},\theta_i\geq0, \Sigma_{i=1}^{p} \alpha_i + \Sigma_{i=1}^{q}\beta_i<1\}
\end{equation}
\section{Necessary tools}
\subsection{Convexity}
W.C. Ip and al. \cite{conv} have proved the convexity of the negative likelihood function in the asymptotic sense for GARCH models. This property allows us the local approximation of this function in the Neighborhood of its minimum by a polynomial of degree two.
\begin{propo}
Suppose $\Theta_1$ is an arbitrary compact, convex subset of $\Theta_0$ and $f^{'}_n(\theta,\varpi)$ the second derivative of $L_n(\theta)$. Then there exist a constant $C>0$ and a set $\Omega$ with $P(\Omega)= 1$ satisfying that for each $\varpi\in\Omega$ and $\theta\in \Theta_1$, there is a positive integer $N(\varpi, \theta)$ such that
\begin{equation}\label{cc}
\upsilon^T f^{'}_n(\theta,\varpi)\upsilon\geq C \upsilon^T \upsilon, \forall n\geq N(\varpi, \theta), \upsilon \in R^{p+q+1}.
\end{equation}
\end{propo}
\subsection{Localisation method}
In this part, we propose a method to search for a block of the form $ \Pi_{i=1}^{p+q+1}]\underline{\theta_i},\overline{\theta_i}[$ containing the maximum sought-after. The method is based on the principle of dichotomy applied for projections of the likelihood function on a  plane with dimension two, we follow the steps
\begin{enumerate}
\item We search a point $\overline{\omega}$ that verifie $ \widehat{L_{1,n}}^{'}(\omega,0.5,...,0.5)>0$, where $\widehat{L_{i,n}}^{'}$ is the $i$ order derivative of the likelihood function with respect to $\omega$.

\item We put $\underline{\theta_i}=0.0001, \forall {i=1, p+q+1}$ and $\overline{\theta_1}=\overline{\omega}, \overline{\theta_i}=0.9999, \forall {i=2, p+q+1}$
\item $\forall {i=1, p+q+1}$\\
\begin{itemize}
  \item if $\widehat{L^{'}_{i,n}}\big(\underline{\theta_1},...,\underline{\theta_i},....,\underline{\theta}_{p+q+1}\big)\widehat{L_{i,n}}^{'}\big(\underline{\theta_1},....,(\underline{\theta}_i+\overline{\theta}_i)/2,....,\underline{\theta}_{p+q+1}\big)<0$ then we replace $\overline{\theta_i}$ by $(\underline{\theta}_i+\overline{\theta}_i)/2$
  \item else if $\widehat{L^{'}_{i,n}}\big(\underline{\theta_1},...,\underline{\theta_i},....,\underline{\theta_{p+q+1}}\big)\widehat{L_{i,n}}^{'}\big(\underline{\theta_1},....,(\underline{\theta}_i+\overline{\theta}_i)/2,....,\underline{\theta}_{p+q+1}\big)>0$ we replace $\underline{\theta}_i$ by $(\underline{\theta}_i+\overline{\theta}_i)/2$
\end{itemize}
\end{enumerate}
 We repeat the step $3$ until $\overline{\theta}_{i}-\underline{\theta}_i\leq 0.05, \forall {i=1, p+q+1}$.
\section{Calculation procedure }
To calculate the maximum likelihood, one passes by the following steps
\begin{enumerate}
\item  Calculate confidence intervals for the unknown parameters using the localisation method.
  \item Make a subdivision of $m$ elements $(\theta_{ij})_{j=1,m}$ for all the confidence intervals $(I_i)_{i=1,p+q+1}$.
  \item Calculate the function values $\big[l_n(\theta_{j})\big]_{j=1,m}$ where $\theta_j=(\theta_{ij})_{i=1,p+q+1}=(\omega_j,\alpha_{1j},...,\alpha_{pj},\beta_{1j},...,\beta_{qj})$, Which is a cut for the curve of the likelihood function on the diagonal plane of the confidence region.
  \item Numerical approximation for the orthogonal projection of the log likelihood function cut by polynomials of order two taking the form $a_0+a_1 \theta_i+a_2 \theta_i^2$: { we use the least squares method}.
  \item Calculate these maximum.
\end{enumerate}

\section{Example of application ARCH(1)}
The $ARCH(1)$ process is presented as solution of the system
\begin{equation}\label{arch}
\left\{
  \begin{array}{ll}
& X_t= \sigma_t \cdot \varepsilon_t,\\
&\sigma_t^2=\omega+\alpha_1 X_{t-i}^2.\\
  \end{array}
\right.
\end{equation}
where $\omega>0$ and $0<\alpha_1<1$.\\
Let  be $(x_t)_{t=1,100}$  the simulated first order conditionally heteroskedastic autoregressive time series $ARCH(1)$ with ($\omega=0.8$ and $\alpha_1=0.3$) presented by the table \ref{Table2}.\\
\begin{table}
\centering
\begin{tabular}{|c|c|c|c|c|c|c|c|c|c|c|}
  \hline
Time $t$ & 1 & 2 & 3 & 4 & 5 & 6 & 7 & 8 & 9 & 10 \\
$x_t$& -1.348 & -0.05 & -0.063 & 2.055 & 0.815 & 1.893 & -2.277 & -1 & 0.782 & 0.351 \\
\hline
Time $t$ &  11 & 12 & 13 & 14 & 15 & 16 & 17 & 18 & 19 & 20 \\
$x_t$&  -0.791 & -2.24 & 1.723 & 0.667 & -0.015 & 0.464 & 0.22 & -0.737 & 0.434 & 0.643 \\
\hline
  Time $t$ &21 & 22 & 23 & 24 & 25 & 26 & 27 & 28 & 29 & 30 \\
 $x_t$& -0.259 & -0.313 & 0.907 & 1.268 & -0.888 & -1.376 & -1.367 & -0.805 & 0.528 & -0.813 \\
\hline
  Time $t$ &31 & 32 & 33 & 34 & 35 & 36 & 37 & 38 & 39 & 40 \\
  $x_t$&-1.89 & -2.051 & 1.94 & 1.643 & -1.071 & -0.336 & 1.085 & -0.766 & 1.59 & 0.993 \\
\hline
Time $t$ &  41 & 42 & 43 & 44 & 45 & 46 & 47 & 48 & 49 & 50 \\
  $x_t$&-1.162 & 2.985 & -0.1 & -0.732 & 0.391 & 0.132 & -2.224 & -0.271 & -0.336 & -1.606 \\
\hline
  Time $t$ &51 & 52 & 53 & 54 & 55 & 56 & 57 & 58 & 59 & 60 \\
  $x_t$&0.509 & -0.026 & 0.468 & -1.626 & 1.219 & 0.315 & -0.416 & 0.636 & 0.848 & -1.011 \\
\hline
  Time $t$ &61 & 62 & 63 & 64 & 65 & 66 & 67 & 68 & 69 & 70 \\
  $x_t$&1.152 & 0.085 & -0.114 & -0.744 & 1.456 & -0.243 & -0.332 & -0.078 & 0.678 & 1.668 \\
\hline
  Time $t$ &71 & 72 & 73 & 74 & 75 & 76 & 77 & 78 & 79 & 80 \\
  $x_t$&-1.499 & -1.347 & -0.886 & -0.578 & -1.94 & 0.156 & -0.082 & -0.173 & -0.63 & -0.677 \\
\hline
  Time $t$ &81 & 82 & 83 & 84 & 85 & 86 & 87 & 88 & 89 & 90 \\
 $x_t$& -0.397 & 1.283 & 0.479 & -1.035 & -0.917 & 1.054 & -0.605 & 0.412 & -1.055 & 0.994 \\
\hline
  Time $t$ &91 & 92 & 93 & 94 & 95 & 96 & 97 & 98 & 99 & 100 \\
$x_t$&  -0.259 & -0.313 & 0.907 & 1.268 & -0.888 & -1.376 & -1.367 & -0.805 & 0.528 & -0.813 \\
\hline
\end{tabular}
\begin{center}
\caption{Table of the time serie $x_t$.}
\label{Table2}
\end{center}
\end{table}
We plot this time series as function of time
\begin{figure}[!t]
\centering
\includegraphics[width=1\textwidth]{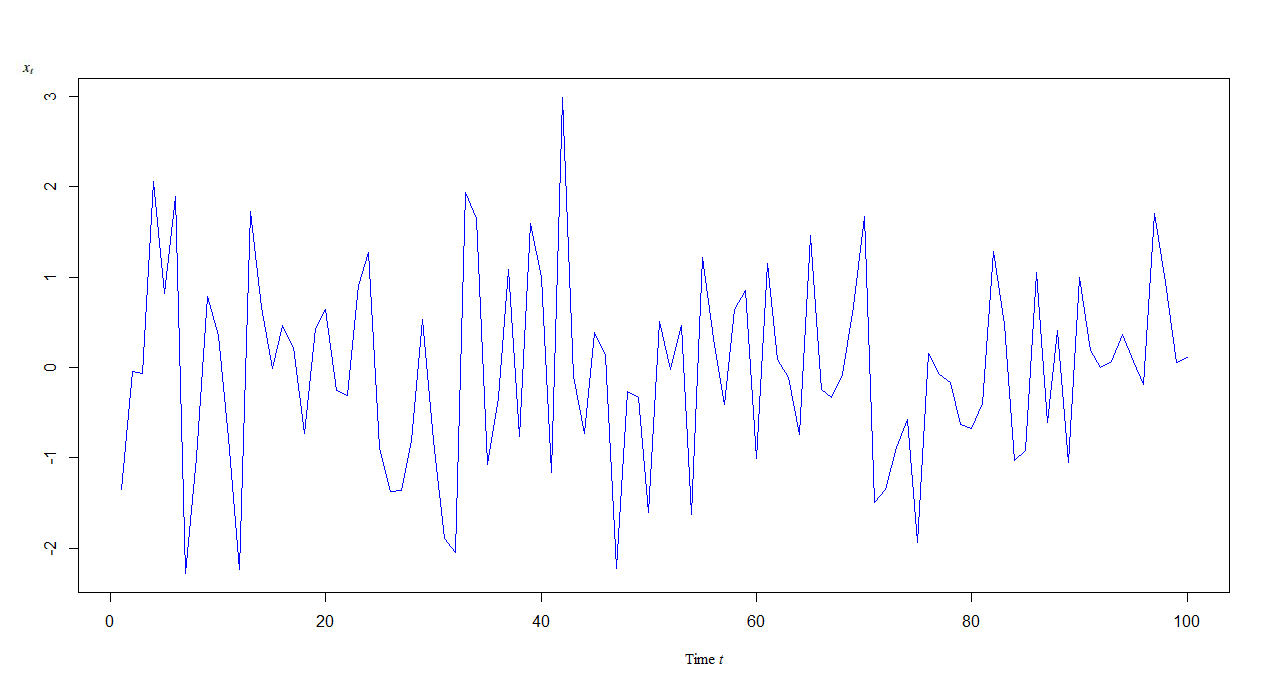}
\caption {The time serie $x_t$.}
\label{fig:fig}
\end{figure}

The log likelihood of an $ARCH(1)$ process is given by
\begin{eqnarray*}
  L_n(\theta)  &=& -\frac{1}{2}\sum^n_{t=1} \Big[log \big(\omega+\alpha_1 X_{t-1}^2\big)^2 + \big(\omega+\alpha_1 X_{t-1}^2\big)^{-2} X_t^2\Big].
\end{eqnarray*}
to obtain $\widehat{L}_n$ we replace $(X_{-i}), i\geq0$ by zero.\\
Now, we proceed to the calculation of the maximum likelihood using our method
\begin{enumerate}
  \item We remark that $ \widehat{L_{n,1}^{'}}(0.8001,0.5)>0$, therefore $\omega\in]0.0001,0.8001[$.
\begin{table}
\centering
\begin{tabular}{|c|c|c|c|c|c|}
  \hline
$w$ &0.0001 & 0.2001 & 0.4001 & 0.6001 & 0.8001 \\
  \hline
$\widehat{L_{n,1}^{'}}(w,0.5)$& -7613853&-455.4789&-93.72643& -19.2947& 5.967303\\
  \hline
\end{tabular}
\begin{center}
\caption{Table of the derivative values.}
\label{Table2}
\end{center}
\end{table}
  \item Using the localisation procedure, we find that the region $] 0.7751000 ,0.8501000[\times]0.2813219, 0.3625687[$ contains the maximum.
  \item We calculate the function values on the diagonal plane of the confidence region, which give the table \ref{Table2}.
\begin{table}
\centering
\begin{tabular}{|c|c|c|c|c|c|c|c|c|c|c|}
 \hline
 $\omega$ & 0.7751 & 0.7759 & 0.7766 & 0.7774 & 0.7781 & 0.7789 & 0.7796 & 0.7804 & 0.7812 & 0.7819 \\
$\alpha_1$ &  0.2813 & 0.2821 & 0.283 & 0.2838 & 0.2846 & 0.2854 & 0.2862 & 0.2871 & 0.2879 & 0.2887 \\
$\widehat{L}_n(\omega,\alpha_1)$  &  109.257 & 109.251 & 109.245 & 109.239 & 109.233 & 109.227 & 109.222 & 109.217 & 109.212 & 109.207 \\
\hline
  $\omega$ &0.7827 & 0.7834 & 0.7842 & 0.7849 & 0.7857 & 0.7865 & 0.7872 & 0.788 & 0.7887 & 0.7895 \\
  $\alpha_1$ &0.2895 & 0.2903 & 0.2912 & 0.292 & 0.2928 & 0.2936 & 0.2945 & 0.2953 & 0.2961 & 0.2969 \\
  $\widehat{L}_n(\omega,\alpha_1)$  &109.202 & 109.198 & 109.194 & 109.19 & 109.186 & 109.182 & 109.179 & 109.176 & 109.173 & 109.17 \\
\hline
  $\omega$ &0.7903 & 0.791 & 0.7918 & 0.7925 & 0.7933 & 0.794 & 0.7948 & 0.7956 & 0.7963 & 0.7971 \\
  $\alpha_1$ &0.2977 & 0.2986 & 0.2994 & 0.3002 & 0.301 & 0.3018 & 0.3027 & 0.3035 & 0.3043 & 0.3051 \\
  $\widehat{L}_n(\omega,\alpha_1)$  &109.168 & 109.165 & 109.163 & 109.161 & 109.159 & 109.158 & 109.156 & 109.155 & 109.154 & 109.153 \\
\hline
  $\omega$ &0.7978 & 0.7986 & 0.7993 & 0.8001 & 0.8009 & 0.8016 & 0.8024 & 0.8031 & 0.8039 & 0.8046 \\
  $\alpha_1$ &0.3059 & 0.3068 & 0.3076 & 0.3084 & 0.3092 & 0.3100 & 0.3109 & 0.3117 & 0.3125 & 0.3133 \\
  $\widehat{L}_n(\omega,\alpha_1)$  &109.153 & 109.152 & 109.152 & 109.152 & 109.152 & 109.152 & 109.153 & 109.153 & 109.154 & 109.155 \\
\hline
  $\omega$ &0.8054 & 0.8062 & 0.8069 & 0.8077 & 0.8084 & 0.8092 & 0.8099 & 0.8107 & 0.8115 & 0.8122 \\
  $\alpha_1$ &0.3141 & 0.315 & 0.3158 & 0.3166 & 0.3174 & 0.3183 & 0.3191 & 0.3199 & 0.3207 & 0.3215 \\
  $\widehat{L}_n(\omega,\alpha_1)$  &109.156 & 109.157 & 109.159 & 109.161 & 109.162 & 109.164 & 109.167 & 109.169 & 109.171 & 109.174 \\
\hline
  $\omega$ &0.813 & 0.8137 & 0.8145 & 0.8153 & 0.816 & 0.8168 & 0.8175 & 0.8183 & 0.819 & 0.8198 \\
  $\alpha_1$ &0.3224 & 0.3232 & 0.324 & 0.3248 & 0.3256 & 0.3265 & 0.3273 & 0.3281 & 0.3289 & 0.3297 \\
  $\widehat{L}_n(\omega,\alpha_1)$  &109.177 & 109.18 & 109.183 & 109.186 & 109.19 & 109.193 & 109.197 & 109.201 & 109.205 & 109.21 \\
\hline
  $\omega$ &0.8206 & 0.8213 & 0.8221 & 0.8228 & 0.8236 & 0.8243 & 0.8251 & 0.8259 & 0.8266 & 0.8274 \\
  $\alpha_1$ &0.3306 & 0.3314 & 0.3322 & 0.333 & 0.3338 & 0.3347 & 0.3355 & 0.3363 & 0.3371 & 0.3379 \\
  $\widehat{L}_n(\omega,\alpha_1)$  &109.214 & 109.218 & 109.223 & 109.228 & 109.233 & 109.238 & 109.244 & 109.249 & 109.255 & 109.26 \\
\hline
  $\omega$ &0.8281 & 0.8289 & 0.8296 & 0.8304 & 0.8312 & 0.8319 & 0.8327 & 0.8334 & 0.8342 & 0.8349 \\
  $\alpha_1$ &0.3388 & 0.3396 & 0.3404 & 0.3412 & 0.3421 & 0.3429 & 0.3437 & 0.3445 & 0.3453 & 0.3462 \\
  $\widehat{L}_n(\omega,\alpha_1)$  &109.266 & 109.272 & 109.279 & 109.285 & 109.291 & 109.298 & 109.305 & 109.312 & 109.319 & 109.326 \\
\hline
  $\omega$ &0.8357 & 0.8365 & 0.8372 & 0.838 & 0.8387 & 0.8395 & 0.8403 & 0.841 & 0.8418 & 0.8425 \\
  $\alpha_1$ &0.347 & 0.3478 & 0.3486 & 0.3494 & 0.3503 & 0.3511 & 0.3519 & 0.3527 & 0.3535 & 0.3544 \\
  $\widehat{L}_n(\omega,\alpha_1)$  &109.333 & 109.341 & 109.349 & 109.356 & 109.364 & 109.372 & 109.38 & 109.389 & 109.397 & 109.406 \\
\hline
$\omega$ &  0.8433 & 0.844 & 0.8448 & 0.8456 & 0.8463 & 0.8471 & 0.8478 & 0.8486 & 0.8493 & 0.8501 \\
  $\alpha_1$ &0.3552 & 0.356 & 0.3568 & 0.3576 & 0.3585 & 0.3593 & 0.3601 & 0.3609 & 0.3617 & 0.3626 \\
  $\widehat{L}_n(\omega,\alpha_1)$  &109.414 & 109.423 & 109.432 & 109.441 & 109.45 & 109.46 & 109.469 & 109.479 & 109.488 & 109.498 \\
\hline
\end{tabular}
\begin{center}
\caption{Table of $(\omega,\alpha_1), \widehat{L}_n(\omega,\alpha_1)$.}
\label{Table2}
\end{center}
\end{table}
  \item Using the least squares method, we calculate approximations of $\big(\omega_i, \widehat{L}_n(\omega_i,\alpha_{1i})\big)_{i=1,100} $ and $\big(\alpha_{1i}, \widehat{L}_n(\omega_i,\alpha_{1i})\big)_{i=1,100} $  the orthogonal projections of the likelihood function curve cut $\big(\omega_i,\alpha_{1i}, \widehat{L}_n(\omega_i,\alpha_{1i})\big)_{i=1,100} $ on the planes $(o,\omega,ln (\omega, \alpha_1))$, $(o,\alpha_1,ln (\omega, \alpha_1))$ respectively.\\
\begin{itemize}
  \item
The approximation function for the likelihood function Curve cut projection on the plane $(o,\omega,ln (\omega, \alpha_1))$ is
\begin{center}
$f(\theta)=  143.7092  \omega^2 -230.1460  \omega +143.7092
$;
\end{center}
the maximum of this function is $230.1460/(2*143.7092)= 0.8007353$.\\
 In the Figure \ref{fig:fig1}, we represent the projection of the cut of the likelihood function on the plane $(o, \omega, ln (\omega, \alpha_1))$ and its approximation

\begin{figure}[!t]
\centering
\includegraphics[width=1\textwidth]{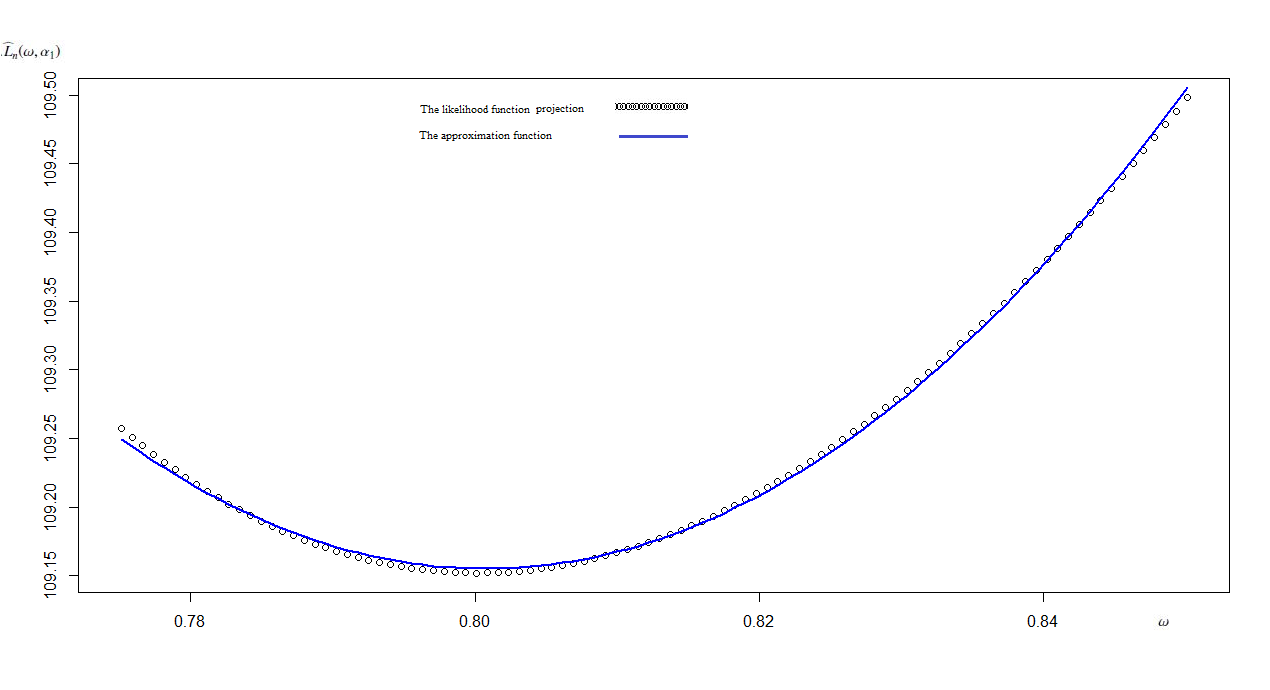}
\caption{The projection of the cut of the likelihood function on the plane $(o, \omega, ln (\omega, \alpha_1))$ and its approximation}
\label{fig:fig1}
\end{figure}

  \item The approximation function for the likelihood function curve cut projection on the plane $(o,\alpha,ln (\omega, \alpha_1))$ is
\begin{center}
$f(\theta)=  120.8546 \alpha_1^2 - 75.7028 \alpha_1 +122.4599$;
\end{center}
the maximum value of this function is $75.7028/(2*120.8546)=0.3090923$.\\

The Figure \ref{fig:fig2} illustrates the projection of the cut of the likelihood function on the plane $(o, \omega, ln (\omega, \alpha_1))$ and its approximation
\begin{figure}[!t]
\centering
\includegraphics[width=1\textwidth]{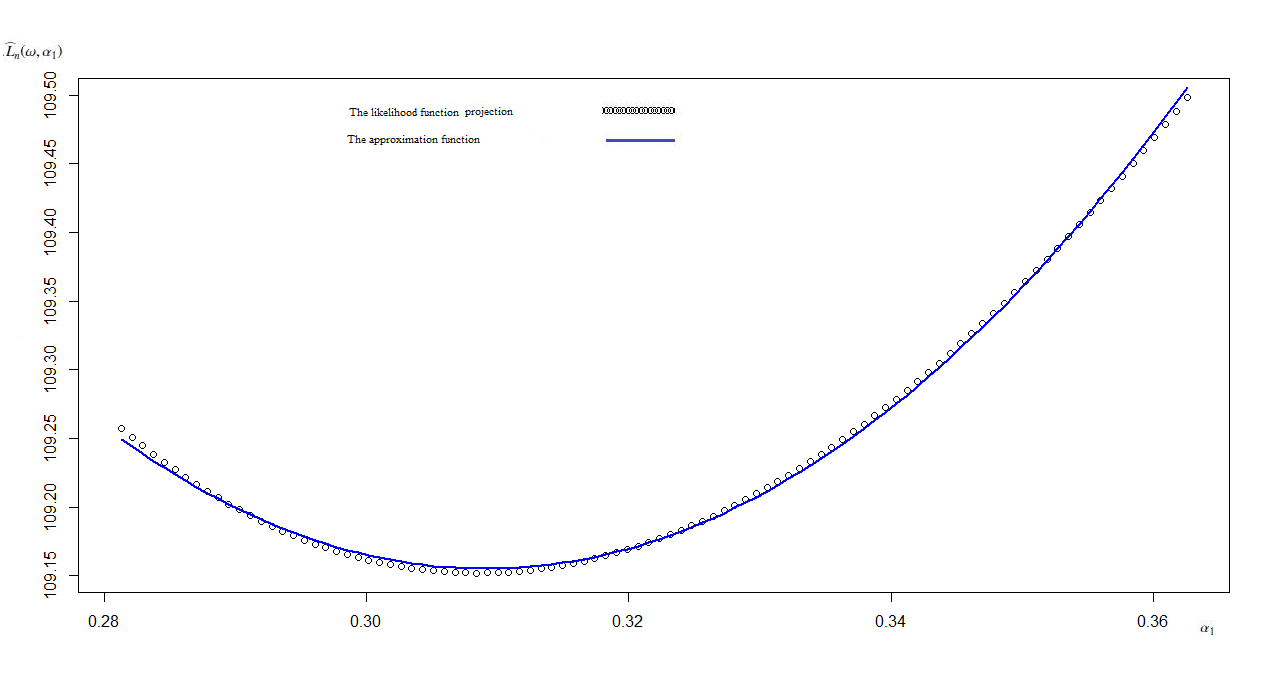}
\caption{The projection of the cut of the likelihood function on the plane $(o,\alpha_1 , ln (\omega, \alpha_1))$ and its approximation}
\label{fig:fig2}
\end{figure}
\end{itemize}
\end{enumerate}
\section{Numerical comparaison study}
To illustrate the performance of the proposed method, we compare it with some popular methods (BGFS, Simplex and OPR) usually used in the Time series parameters calculation. We applied these methods to $1000$ independent replications of two  ARCH$(1)$ process (one with $\omega=1.2$, $\alpha= 0.6$ end the other with $\omega=0.7$, $\alpha= 0.4$ for different sample size $n=100$, $n=200$ and $n=300$, thereafter we compute the root-mean-square error (RMSE) $RMSE (\omega)+RMSE (\alpha_1)$ of the resulting estimation, the results are presented in Table \ref{Table11}.\\

\begin{table}
\centering
\begin{tabular}{ll|c|c|c|c||c|c|c|c|}
\cline{3-3}\cline{4-4}\cline{5-5}\cline{6-6}\cline{7-7}\cline{8-8}\cline{9-9}\cline{10-10}
  & &\multicolumn{4}{c||}{$\omega= 1.2,\alpha_1=0.6$} & \multicolumn{4}{c|}{$\omega= 0.7,\alpha_1=0.4$} \\
\hline
  &Sample size& {Our Method} & {BGFS} &{{Simplex}} & {{OPR}}& {Our Method} & {BGFS} &{{Simplex}} & {{OPR}}\\
  \hline
   &100     &  0.485  &    0.607 &   0.609 &  0.609 &  0.798 &   0.804 &  0.807&0.807\\
   &200 &  0.345 & 0.368   & 0.368  & 0.368 &  0.747 &0.752 &0.752   &0.752 \\
   &300  &  0.292 &0.300    &0.300   & 0.300 &  0.742 & 0.743&0.743   &0.743 \\
\hline \\
\end{tabular}
\begin{center}
\caption{Root Mean Square Error of our method, BGFS, Simplex and OPR for an ARCH$(1)$ processes.}
\label{Table11}
\end{center}
\end{table}
{\bf Conclusion of the numerical comparaison results:} On the one hand, it is clear that the RMSE decreases as the sample size increases, which validates the theoretical results (consistency of the estimators). On the other hand, Table \ref{Table11} show that our method provides more accurate estimation than the BGFS, Simplex and OPR methods.

\end{document}